\documentclass{elsart}
\usepackage{graphicx,latexsym}

\begin{document}
\begin{frontmatter}
\title{Statistical characterization of the fixed income market efficiency} 
\author{Massimo Bernaschi\thanksref{mail}}, 
\author{Luca Grilli},
\author{Livio Marangio}, 
\author{Sauro Succi}
\address{``Istituto Applicazioni del Calcolo'', CNR \\ 
          Viale del Policlinico 137, I-00161 Rome, Italy}
\author{Davide Vergni}
\address{Physics Department and INFM, University of Rome ''La Sapienza''\\ 
P.le Aldo Moro 2, I-00185 Rome, Italy}
\thanks[mail]{Corresponding author, e-mail: massimo@iac.rm.cnr.it}
\begin{abstract}
We present cross and time series analysis of price fluctuations in the
U.S. Treasury fixed income market. By means of techniques borrowed from
statistical physics we show that the correlation among bonds depends
strongly on the maturity and bonds' price increments do not fulfill the
random walk hyphoteses.
\end{abstract}

\end{frontmatter}

The individual evolution of complex systems can be tantalizingly
difficult to describe. However, when appropriate statistical averages
are taken, some form of structural regularity may be observed, that
can be used to model the behavior of the system.

It so happens that the same mathematical techniques are applied to
analyze widely disparate phenomena, such as fluid turbulence and the
price fluctuations in the financial markets. Recently, for instance,
typical concepts of statistical mechanics, such as scaling,
(multi)-fractality, percolation and others are
starting to play a significant role in the quantitative analysis of
the behavior of financial markets.

Actually, fixed-income (FI) markets have received less attention than
stock markets in terms of a quantitative characterization as a
stochastic process. On the other hand, FI markets (specifically US
Treasury securities markets, which is the one we are interested in)
are much less volatile than stock markets since their own
characteristics, in a strong-economics country with low inflation,
prevent the occurrence of wild fluctuations.

Fixed income markets are expected to honor the so called ``one-price
law'' stating that two portfolios of fixed-income securities that
guarantee the investor the same cash-flows and give her the same
future liabilities, must sell for the same price. If any violation of
that constraint should occur, then a profitable and {\em risk-less}
investment opportunity, called an {\em arbitrage}, would arise. But in
an efficient market \cite{Fama} such an opportunity can not last long,
because as investors detect and take advantage of it, a price change
occurs which re-absorbs the initial anomaly.

Note that, although not limited to FI securities (it represents, for
instance the basis of the well-known Black and Scholes option pricing
model), the one-price law is somehow more ``natural'' for such assets
since the market is much more homogeneous and (for all practical
purposes) default risk-free.

It is apparent that, as a consequence of the one-price law, we can
expect strong correlations among bonds. In particular when a
replicating portfolio exists. By ``replicating portfolio'' of a given
bond, we mean a portfolio of bonds which assures the same cash-flow of
the original one.

However, even if the price variations of the bonds are strongly
correlated, so that we can consider the bond dynamics as a collective
motion, the process of price evolution is not known.

The statistical characterization of such process, and the study of the
link between the bonds is by no means of negligible importance, given
that FI markets exceed stock markets in terms of volume liquidity.
This is one of the motivations of this Letter.

We have analyzed 100 US Treasury notes and bonds in the
period between 30/01/97 and 28/09/99 (694 daily
prices). In~\cite{Marangio} we focused on ``cross-section'' data (i.e.,
for a given day, the set of the prices of all Treasury securities
outstanding that day) looking for arbitrage opportunities. We found
that in the range of maturities 5-7 years there was a difference of
(about) one percent in the price of a bond and its replicating
portfolio. For longer maturities (in the range 7-30 years) it was not
possible to build a replicating portfolio so it was meaningless to test
the existence of arbitrage opportunities.

The aim of the present work is twofold: to investigate the consequence
of the one-price law on the {\em cross} correlations and to test the
random walk hyphoteses for the time evolution of the price
fluctuations.

As a starting point, we build the correlation matrix and then classify
the bonds according to a suitable metric.

The correlations between the price dynamics of different assets is of
paramount importance in the analysis of financial markets. For
instance for building multi-factor pricing models it is necessary to
verify some properties of the time-series of returns.

In our case the assets are a set of FI securities $B_i, \;i=1 \ldots N$
($N=100$) and the correlation matrix elements are defined as:
$$C_{ij}=\frac{\langle r_i(t)r_j(t) \rangle - \langle r_i(t) \rangle \langle r_j(t) \rangle}
              {\sqrt{(\langle r_i(t)^2 \rangle - \langle r_i(t) \rangle^2)
                     (\langle r_j(t)^2 \rangle - \langle r_j(t) \rangle^2)}}\,\,,$$
where $r_i(t) \equiv \ln (p_i(t) / p_i(t-1))$ is the return, that is
the logarithmic price change of the $i$-th asset between $t-1$ and
$t$, and $\langle f(t) \rangle$ denotes the time average $1/T \cdot
\sum_{t=1}^T f(t)$ (T is the sample length).

By definition $C_{ij}=0$ if bonds $i$ and $j$ are totally
uncorrelated, whereas $C_{ij}=\pm 1$ in case of perfect
correlation/anti-correlation.

$C$ is a $N \times N$ symmetric matrix with $1$ in the diagonal.  In
our data set $C_{ij}$ is positive for each pair of assets. This means
that, on average, the price increments of U.S. Treasury securities
have the same sign. This is not surprising since the prices of FI
securities issued by a single large issuer like the US Department of
Treasury depend on the same macroeconomics factors like the market
expectations on future base interest rate.

The $T$ values of the logarithmic price increments can be interpreted
as the coordinates in a ${\mathrm {I \!  R}}^T$ space . A ``natural''
Euclidean distance exists for such space, however, following
~\cite{Mantegna}, $C_{ij}$ can be used to define a different metric by
means of the following simple formula:
\begin{equation}
 d(i,j)=\sqrt{2(1-C_{ij})}\,\,. 
 \label{distance}
\end{equation}

This metric has the advantage that just the correlation between assets
matters. The ``absolute'' distance of two assets becomes irrelevant as
it is from a financial viewpoint.

Function (\ref{distance}) fulfills all the three axioms of a distance:
$d(i,j)=0 \Leftrightarrow i=j$, $d(i,j)=d(j,i)$ and $d(i,j) \le
d(i,k)+d(k,j)$ when restricted to the set of time series of assets
with zero mean and variance equal to one.
The theoretical range of the distance is $[0,2]$, but, since
$C_{ij}>0$ for our data, we have $0 \leq \delta \leq \sqrt{2}$.

Equipped with this distance we study the topological arrangement of
the bonds. However we did not follow the approach of \cite{Mantegna}
that starts from the minimum spanning tree connecting the assets of
the portfolio.

\begin{figure}
\begin{center}
\includegraphics[width=0.8\textwidth,height=0.3\textheight]{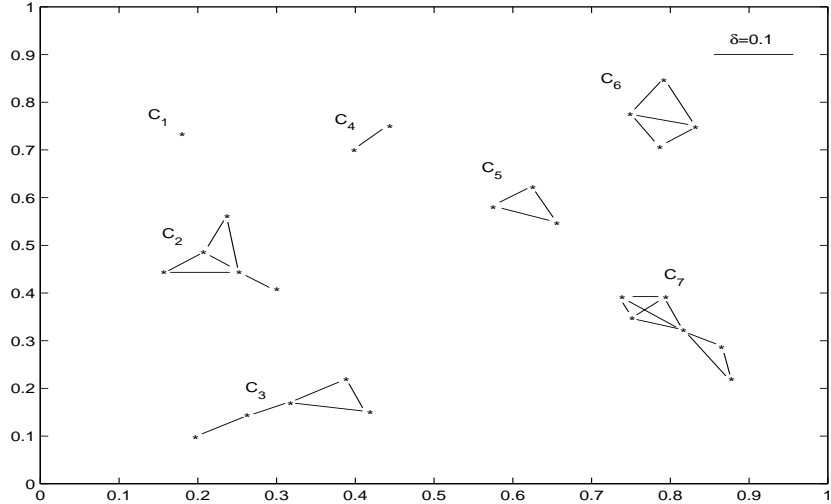}
\caption{A simplified representation of the invasion
         percolation algorithm. We show, in a 2D space, some
         points (i.e. bonds, $B_i$) connected by links whose length is less
         than $\delta$. Each connected group forms a cluster, $C_i$.}
\label{clufig}
\end{center}
\end{figure}

By means of an invasion-percolation algorithm, which groups together 
bonds whose distance falls within a threshold $\delta$,
we analyze the structure of the clusters as a function of $\delta$.
A bond $b_i$ belongs to a cluster if there is at least one bond
of that cluster whose distance from $b_i$ is less than $\delta$
(see Figure~\ref{clufig}).

With this choice we have that the minimal distance between
two clusters is greater than $\delta$.

The number of clusters can vary between $N$ disconnected clusters of
size $1$ (by size we mean the number of bonds in the cluster) when
$\delta = 0$, and $1$ large cluster of size $N$ when $\delta = \sqrt{2}$.
According to standard percolation-theory \cite{Stauffer} a critical
percolation threshold $0<\delta_c<\sqrt{2}$ is expected: $\delta_c$ is the
smallest distance for which only $1$ cluster appears.

This technique allows to gain information about the topological
structure of the bonds' space using the distance (\ref{distance}).
Other methods, which do not depend on a distance parameter, like $\delta$,
are less effective than our invasion-percolation tool.

\begin{figure}
\begin{center}
\includegraphics[width=0.8\textwidth,height=0.3\textheight]{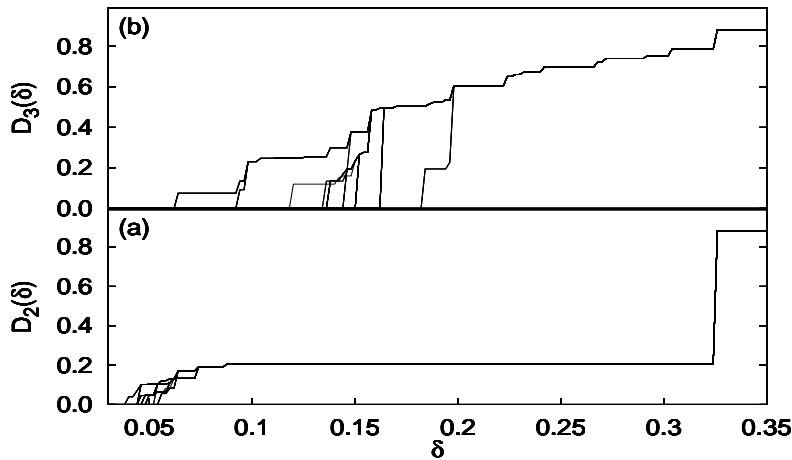}
\caption{Clusters formation. The top half (b) describes the $C_3$ cluster
whereas the bottom half (a) refers to $C_2$. In each half, the upper
continuous line represents the evolution of the clusters'
diameter. Each other line represents the evolution of single bonds (or
small cluster) which are absorbed by the main cluster. We have shown
only ten bonds for clusters (which are the same as Figure 3). 
The collapse of two lines implies the merge of the corresponding clusters.}
\label{cluster}
\end{center}
\end{figure}

\begin{figure}
\begin{center}
\includegraphics[width=0.8\textwidth,height=0.3\textheight]{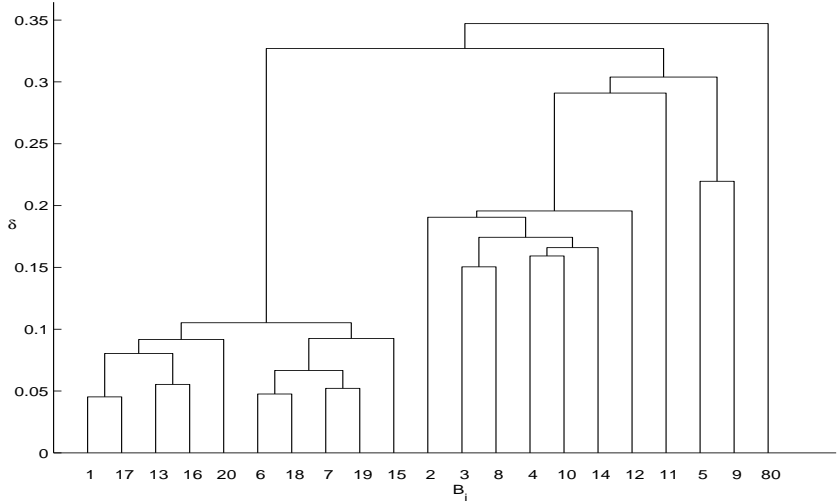}
\caption{Dendrogram representing the cluster structure of the bonds.
         Each horizontal line measures at which $\delta$ we have a fusion
         between two bonds, a bond and a cluster or two clusters.}
\label{fig:dendro}
\end{center}
\end{figure}

We found that there are essentially three distinct clusters
$C_1$,$C_2$ and $C_3$, up to the critical percolation threshold
$\delta_c \sim 0.347$:
\begin{trivlist}
\item[$C_1$] is a very small cluster consisting of just a single
bond, which is absorbed by the merge of $C_2$ and $C_3$
at the percolation threshold $\delta_c \sim 0.347$
(in the Figure~\ref{fig:dendro} $C_1$ is the bond 80).

\item[$C_2$] is a medium-size very compact cluster. It consists of
$29$ bonds which groups together when $\delta \sim 0.087$. This cluster does not
change up to $\delta < 0.325$, when it finally merges with $C_3$.

\item[$C_3$] is the largest cluster which forms roughly at $\delta \sim
0.157$ and absorbs a variety of micro-clusters up to $\delta < 0.305$ as well
as $C_2$ at $\delta > 0.325$.
\end{trivlist}
In Figure~\ref{cluster} we show the diameter of the clusters, $D_c(\delta)$ 
at varying $\delta$. We define 
\begin{equation}
   D_c(\delta) \equiv \max_{i,j\in C_c} d(i,j) \,\,.
   \label{diameter}
\end{equation}
The diameter gives information about the spread of a cluster.
From the Figure~\ref{cluster} one can see the clusters formation
of $C_2$ and $C_3$.
An alternative representation of the cluster structure is offered by
the graph in Figure \ref{fig:dendro} that consists of lines (having
the shape of an ``upside-down U'') connecting bonds in a hierarchical
tree built by means of a linkage algorithm \cite{Griffiths}. 
The height of each U line is the distance between
two clusters. In Figure \ref{fig:dendro}, for the sake of clarity,
just a subset of bonds is included in the {\em dendrogram} (that is
the name of this particular graph).
Besides the existence of three main classes of bonds, the dendrogram
makes easier to locate the ``distance'' between these groups. We
recall that such a distance has a financial meaning since it measures
the correlation among bonds and bond classes.
Taking into account the features of the FI securities which form the
three clusters, some additional considerations can be made.

No FI security with maturity greater than 12 years is found in $C_1$
and $C_3$ whereas the second cluster, $C_2$, is composed exclusively
by such ``long term'' bonds.
It is worthy to note that for this class of bonds no replicating
portfolio can be determined meaning that direct test of the one-price
law is not possible. Nevertheless, $C_2$ is very compact (the maximum
diameter $D$ is equal to 0.2077 ) meaning that the correlation among
the corresponding FI securities is very strong.

Usually only large risk-adverse investors (like insurance companies
or mutual funds) are interested in long term bonds (whose maturity can
be up to thirty-years for US Treasury issues), so a possible
explanation is that their expectations (that take into account a
number of macroeconomics factors) about the market evolution are so
similar that the behavior of long term bonds prices does not reflect
any difference in the {\em perceived value} of such assets.

Third cluster collects all the other securities, which are the various
short and medium term obligations. For many of these assets it is
possible to find a replicating portfolio that fulfills the one-price
law. However, the profile of the people interested in these assets is
much more variegated so the investors may act in many different
ways. Such situation generates a minor correlation among the assets,
that is, on average, a larger distance. This is reflected in the
structure of $C_3$ that is wider than $C_2$.

So it looks like that, in spite of its clear and unambiguous
definition, the correlations imposed by the one-price law are weaker
than those produced by ``common expectations'' of few large investors.

From the theoretical point of view a non perfect correlation among the
bonds returns implies that ``classical'' single factor models for the
term structure of interest rate like those by Vasicek \cite{Vasicek}
or Cox, Ingersoll and Ross \cite{Cox} can not explain the complex
behavior of empirical data.

Besides the correlations among the FI securities, we have analyzed the
properties of bond price dynamics. The price fluctuations of a single
bond may give an insight into the problem of FI markets
efficiency~\cite{Fama}. If price increments are uncorrelated
variables, with zero mean value, it is not possible to extract any
information about the future evolution of an asset price by looking at
its fluctuations in the past (``weak'' efficiency).

From Bachelier~\cite{Bachelier} to nowadays many attempts have been
performed to model the statistical properties of financial time
series. The distribution function for the returns was considered for a
long time a zero mean Gaussian. Mandelbrot proposed, in
1963~\cite{MandelbrotII}, a Levy distribution for $r_t$. Recently
Mantegna and Stanley \cite{MantegnaStanley} found that a truncate Levy
distribution fits very well the experimental data.  Regardless of the
specific distribution function, all these models assume that the
returns are independent random variables.  When the increments $r_t$
are, at least, uncorrelated (that is a weaker condition compared to
the independence) a model is considered belonging to the the class of
``random walks''.

To obtain information about the probability distribution of the price
increments we study the moments of

\begin{equation}
   r_t(\tau) = \sum_{t'=1}^\tau r_{t+t'} = \ln \frac{P_{t+\tau}}{P_t}\,\,,
   \label{sumret}
\end{equation}

The Structure Function \cite{Frisch} is widely used in fully developed
turbulence to study the fluctuations of the velocity field of a
fluid. In our case, the study of the moments of $r_t(\tau)$ is
equivalent to the study of the structure function of the logarithmic
price:

\begin{equation}
   S_q(\tau) \equiv \langle |r_t(\tau)|^q \rangle_t
                =   \langle |\ln P_{t+\tau} - \ln P_t|^q \rangle_t\,\,.
   \label{strfun}
\end{equation}

The theory of stable distributions \cite{Kolmo} states that, in case
of independent variables, the moments behave as $S_q(\tau) \sim \tau^{hq}$, 
where $h$ is the characteristic exponent of the process.
For Gaussian increments (actually, in all cases where the central
limit theorem applies) $h$ is equal to $1/2$.

We have computed $S_q(\tau)$ for bonds belonging to each cluster
described above and we found that $S_q(\tau) \sim \tau^{\zeta(q)}$
where $\zeta(q)$ is a non linear function. The exponents have been
evaluated by means of the Extended-Self-Similarity technique, which
consists in log-plotting the generic $p^{th}$ order structure function
$S_p$ versus $S_2$ instead of $\tau$ \cite{ESS}. One of the advantages
of the ESS is that it offers a higher level of statistical confidence
since it minimizes finite-size effects.
\begin{figure}
\begin{center}
\includegraphics[width=0.8\textwidth,height=0.3\textheight]{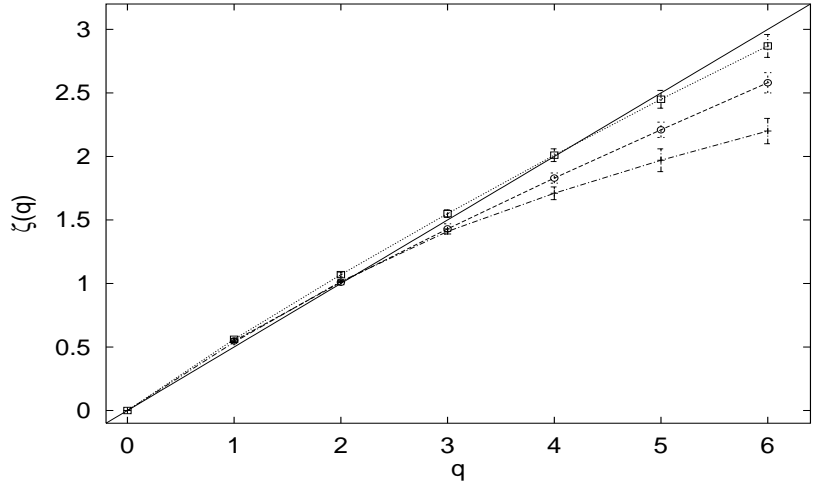}
\caption{$\zeta(q)$ versus $q$ for representative bonds belonging to
the three main clusters: $C_1$ ($\Box$), $C_2$ (+) and $C_3$
($\circ$).}
\label{fig:scaling}
\end{center}
\end{figure}
The results reported in Figure \ref{fig:scaling} provide solid evidence
of a multi-fractal behavior on top of a random walk, $h=1/2$, for
clusters $C_2$ and $C_3$.
All the bonds belonging to the cluster $C_2$ show the same scaling
exponents (variations are smaller than few part-per-thousand), instead
for the cluster $C_3$ the scaling behaviour is more variable, but it
always shows a non linear $\zeta(q)$.

It is apparent that the bond which is less correlated with others,
that is the single bond belonging to $C_1$ has a scaling behaviour
closer to that expected for a random walk.

The non-linearity of $\zeta(q)$ indicates that the $r_t$ are not
independent. This means that the random walk hypothesis for the price
increments in FI markets does not apply. Similar analyses based on the
the structure function have been performed for currency and equity
markets~\cite{Baviera,Brachet}. In both cases the signals show a multi-fractal behavior,
which definitely brings toward models with non independent price increments.

From a statistical point of view, the multifractality of a signal
implies different probability distribution functions of the
$r_t(\tau)$ at different scales $\tau$. In financial terms this means
that price fluctuations of different magnitude behave in a non-uniform
way. For instance, the clear concave-shape of $\zeta(q)$ for large $q$
in Figure \ref{fig:scaling} indicates that large price fluctuations
have an anti-persistent behavior (specially for the bonds in $C2$).

Recently, it has been pointed out by \cite{Bouchaud} that it is
possible to get an apparent multifractal behaviour due to strong time
correlations of returns. In our data set, however, the most interesting 
point is that bonds within a cluster behave in a homogeneous way whereas 
there are remarkable differences between the clusters.

We like to stress that the failure of the random walk hypotheses does
not entail that it is easy to develop ``strategies'' for taking
advantage of the correlations present in the bonds' price increments.
At this time, the only evidence is that conventional ARCH (Auto
Regressive Conditional Heteroskedasticity) models are incompatible
with the scaling properties of price fluctuations.

By now, financial markets analysts have started to accept that the
fine structure of the securities markets and frictions in the trading
process can generate a certain degree of predictability
\cite{Campbell}. However, much work remains to be done because no
existing model is able to explain the behavior of price increments
across a range of different time scales. Maybe, to reach this goal, it
is necessary to model the behavior of the agents, people who buy and
sell assets. In \cite{LeBaron} the properties of the time series in an
artificial stock market have been studied. We expect to perform a 
similar experiment for the fixed income market.

We thank INA-SGR for providing us with the data used in this work and
Dr. Alberto Cybo-Ottone for useful discussions.

\end{document}